\documentstyle[11pt,aaspp4]{article}                 % <=== 1 column
\lefthead{Treves \& Turolla}
\righthead{VACUUM BREAKDOWN NEAR A BLACK HOLE}
\begin{document}

\title{Vacuum Breakdown near a Black Hole Charged \\
by Hypercritical Accretion}
\author{Aldo Treves}
\affil{II Faculty of Sciences, University of Milano, \\ Via Lucini 3,
22100, Como, Italy \\ e--mail: treves@uni.mi.astro.it}
\and
\author{Roberto Turolla}
\affil{Dept. of Physics, University of Padova, \\ Via Marzolo 8, 35131 Padova,
Italy \\ e--mail: turolla@pd.infn.it}

\begin{abstract}

We consider a black hole accreting spherically from the surrounding medium. 
If accretion produces a luminosity close to the Eddington limit the hole 
acquires a net charge so that electrons and ions can fall
with the same velocity. The condition for the electrostatic field to be
large enough to break the vacuum near the hole
horizon translates into an upper limit for the hole mass, $M\sim 6.6\times 
10^{20}\, {\rm g}\,.$ The astrophysical conditions under which this
phaenomenon can take  place are rather extreme, but in principle 
they could be met by a mini black hole residing at the center of a star.

\end{abstract}

 
\section{Charging a Black Hole by Accretion}

Let us consider a star of mass $M$ which undergoes spherical accretion
and assume, for simplicity, that the accreting material is ionized hydrogen.
If the accretion luminosity is $L$, the infalling electrons experience
a radiative force
\begin{equation}
\label{frad}
F_R = {{\sigma_TL}\over{4\pi cr^2}}
\end{equation}
where $\sigma_T$ is the Thomson cross section.
Since the radiation drag acting on the protons is a factor $(m_e/m_p)^2$ 
smaller,
electrons and protons are subject to different accelerations, so
an electrostatic field is built up at the star surface in order to 
equalize the two flows. As first
discussed by Shvartsman (1970\markcite{sh70}, see also Michel 
1972\markcite{mi72}; Maraschi, Reina, \&
Treves 1974\markcite{mrt74}, 1978\markcite{mrt78}; Turolla, {\it et al.\/} 
1997\markcite{tu97}) the positive charge $Q$
associated with the electrostatic field is approximately given by
\begin{equation}
\label{char}
Q = {{GMm_p}\over e}{L\over{L_E}}
\end{equation}
where $L_E = 4\pi GMm_pc/\sigma_T$ is the Eddington luminosity. Note that
in the critical case ($L= L_E$)
the charge to mass ratio is independent of the mass. If the spherical
star is replaced by a non--rotating black hole the above considerations hold 
the same. By analyzing the equation
of motion of a test particle in the Reissner--N\"ordstrom spacetime subject 
to the
drag exerted by a radiation field of luminosity $L$, it is easy to see
that the required charge is still given by equation (\ref{char}), where now $L$ 
is the luminosity measured by an observer comoving with the particles
(see Appendix). We note that close to the black hole matter flows in
supersonically so the ballistic appoximation used to study the motion of
the particles is justified. Besides, for the black hole masses we are
interested in (see the following section) the characteristic time for
Coulomb collisions between electrons and protons (see e.g. Spitzer
1956\markcite{spi56}, pag. 80) is at least one order of
magnitude larger that the free--fall time and the two species are
decoupled.  
Although formally the spacetime is Reissner--N\"ordstrom, deviations from the
Schwarzschild metric are not expected to be of any relevance since the
black hole charge (in geometrized units) is $Q\lesssim (m_p/e)M\ll M$
(see Michel 1972\markcite{mi72}). This is in agreement with the general 
argument by Wald
(1984\markcite{wa84}, see also Novikov, \& Frolov 1989\markcite{nf89}) that 
the charge to mass ratio of
a black hole must be smaller than $m_p/e$ in  any realistic situation.

There is  an important difference between the case of a classical star and
a black hole as far as the accretion luminosity is concerned. For accretion
onto a star of radius $R_*$, the efficiency, assuming Newtonian gravity,
is simply $\eta = GM/R_*c^2$, while in the case of spherical accretion
onto a black hole, $\eta$ depends substatially on many factors, like
the efficiency of radiative processes, or the strength of the frozen--in
magnetic field (see e.g. Shapiro 1973a\markcite{sh73a}, b\markcite{sh73b}; 
M\`esz\`aros 1975\markcite{me75}) and is typically much lower.

\section{Particle Creation}

The issue of pair creation in the electric field of a charged black hole was 
investigated by many authors starting from the early '70s (see 
e.g. Novikov, \& Frolov 1989 and references therein). The main conclusion 
is that the hole charge is radiated away over a timescale shorter than
the characteristic time for the hole to evaporate because of Hawking radiation.
This result, however, holds if the black hole has a ``seed'' charge which
is not resupplied as time elapses. In the following we will give a rough 
estimate of the conditions
under which the accretion--induced electrostatic field $E$ is likely to
produce electron--positron
pairs. We neglect all curvature effects and
assume that the condition for vacuum breakdown (Klein instability) 
is given by the
semi--classical expression (see e.g. Novikov, \& Frolov 1989\markcite{nf89}, 
p. 190)
\begin{equation}
\label{break}
eE\lambda_C=2m_ec^2
\end{equation}
where $\lambda_C$ is the Compton wavelength of the electron. On a scale of 
the order of the Schwarzschild radius $r_G$ and assuming $L\sim L_E$, the 
electrostatic field is
\begin{equation}
\label{efield}
E\sim {Q\over{r_G^2}}\sim {{m_pc^4}\over{4GMe}}\,.
\end{equation}
Following Gibbons (1975)\markcite{gi75}, we take as a first estimate of 
the breakdown
condition what results inserting equation (\ref{efield}) into equation 
(\ref{break})
\begin{equation}
\label{condit}
{{hm_pc^3}\over{4GMm_e}}=2m_ec^2\,.
\end{equation}
Equation (\ref{condit}) translates into an upper limit for the black hole mass, 
above which the electric field is too weak to give raise to pairs
\begin{equation}
\label{mcrit}
M\lesssim {1\over 8}{{m_p}\over{m_e^2}}{{hc}\over G}=M_c\sim 6.6\times 
10^{20}\, {\rm g}\,.
\end{equation}
Note that $(hc/G)^{1/2}=5.5\times 10^{-5}\, {\rm g}$  is the Planck mass.

A black hole of mass $\approx 10^{20}\, {\rm g}$  has a gravitational radius 
$r_G\approx 10^{-8}\, {\rm cm}$ and it is a 
rather peculiar object. Obviously black holes of such a low mass 
can not be produced by stellar collapse and one can only appeal to a 
cosmological origin (see e.g. Carr, \& Hawking 1974\markcite{ch74}).  
Since $M_c$ is substantially larger than the 
mass below which Hawking radiation evaporates the holes in a Hubble time, 
$M_H\sim 5\times 10^{14}\, {\rm g}$, one can argue that 
if mini black holes of $\approx 10^{20}\, {\rm g}$ were created
in the initial phases of the Universe, they should still survive.

The condition for producing an electric field which breaks the vacuum was
derived assuming that $L\sim L_E\sim 7\times 10^{24}\, {\rm erg\, s}^{-1}$
for $M\sim 10^{20}\, {\rm g}$. Assuming an extreme efficiency
$\eta =0.1$, this corresponds to an accretion rate $\dot M\sim 8\times 10^4\, 
{\rm g\, s}^{-1}$. If the Bondi--Hoyle accretion scenario applies, 
the accretion radius is
\begin{equation}
\label{racc}
R_a = {{2GM}\over{v^2}}
\end{equation}
where $v$ is the relevant velocity. The accretion rate turns out to be
\begin{equation}
\label{mdot}
\dot M = 4\pi R_a^2\rho v\, .
\end{equation}

For $v\sim 4\times 10^7\, {\rm cm\, s}^{-1}$ a typical value for the thermal
velocity at the center of low main sequence stars, this implies 
$R_a\sim 10^{-2}\, {\rm cm}$ and a density 
$\rho\sim 3\, {\rm g\,cm}^{-3}$. The time scale for the hole mass 
to change is  $\sim M/\dot M \sim 4\times 10^7$ yr.

The conditions for critical accretion can be found at centers 
of stars, and, possibly, in a cosmological scenario as well.
If pair production due to vacuum breakdown does occur, one should take into 
account a possible reduction of the Eddington luminosity, if the atmosphere
surrounding the mini black hole is dominated  by pairs. This in turn may 
put the system in a condition below the threshold for vacuum breakdown. The 
situation is analogous to that considered by us in the contest of hot 
solution in accreting neutron stars yielding pair production (Zane, Turolla,
\& Treves 1998\markcite{ztt98}). 
Still the basic point which one should bear in mind is that the critical 
luminosity for these mini black holes  would be orders of magnitude 
below that of the host star. Unless a star contains billions of mini holes, no 
observational consequence can come to our mind.

\section{Conclusion}

Using semiclassical arguments we have shown that a black hole of mass
$\lesssim 10^{20}\, {\rm  g}$ accreting hypercritically may produce an electric 
field which is strong enough to give raise to pair production via 
the Klein instability. The astrophysical scenarios in which these conditions
are met are clearly extreme, but not totally unconceivable.
The main merit of our argument may be that using a very simple and easily 
understood mechanism for charging a black hole, it is found that vacuum 
breakdown occurs for masses which are three orders of magnitude larger than
the limiting mass at which the time for evaporation due to Hawking
radiation compares with the age 
of the Universe. In concluding we would like to state clearly that the 
process of particle creation envisaged here is rather different from 
that proposed by Hawking (1974)\markcite{ha74} which is a pure 
curvature effect. Here pair production is due only to the presence of 
an electric field
outside the horizon and the field is a direct consequence of accretion.

\acknowledgments
We are grateful to  Maurizio Martellini for enlightening discussions.

\appendix
\section{Appendix}

In a Reissner--N\"ordstrom spacetime 

\begin{equation}
\label{rn}
ds^2 = -\left(1 - {{2M}\over r} + {{Q^2}\over {r^2}}\right)\,dt^2 +
\left(1 - {{2M}\over r} + {{Q^2}\over {r^2}}\right)^{-1}\,dr^2+r^2d\Omega^2\,,
\end{equation}
the equations of motion of a proton (p) and of an electron (e) are written as

\begin{equation}
\label{aprot}
a^i_p = {e\over{m_p}}F^i_ju^j_p
\end{equation}
\begin{equation}
\label{aelec}
a^i_e = -{e\over{m_e}}F^i_ju^j_e + f^i
\end{equation}
where $u^i$ and $a^i$ are the particle 4--velocity and 4--acceleration, $F^i_j$
is the electromagnetic tensor and $f^i$ is the radiative 4--force per 
unit mass (see e.g. Ruffini 1973\markcite{ru73}). Introducing a tetrad 
field $e^i_{\hat a}$ comoving with the electrons (see e.g.
Nobili, Turolla, \& Zampieri 1991\markcite{ntz91}), 
$a_{\hat r}=e^i_{\hat r}a_i$ gives the
particle acceleration measured by the comoving observer. The request that
electrons and protons fall with the same speed translates then into
\begin{equation}
\label{eqacc}
-{e\over{m_e}}F_{ij}e^i_{\hat r}u^j_e + f_{\hat
r}={e\over{m_p}}F_{ij}
e^i_{\hat r}u^j_p
\end{equation}
which is equivalent to
\begin{equation}
\label{frhat}
f_{\hat r}\simeq {e\over{m_p}}F_{ij}e^i_{\hat r}u^j_p
\end{equation}
where 
\begin{equation}
\label{frhatdef}
f_{\hat r}={{\sigma_TL}\over{4\pi r^2cm_e}}
\end{equation}
is the radiative force (per unit mass) in the electron
rest frame. Since the only non--vanishing components of the EM tensor
are
\begin{equation}
\label{emtens}
F_{01}=-F_{10}= -{Q\over{r^2}}\,,
\end{equation}
equation (\ref{frhat}) reduces to
\begin{equation}
{{\sigma_TL}\over{4\pi r^2cm_e}}\simeq {e\over{m_e}}{{\gamma^2F_{01}}\over
{\sqrt{-g_{00}}\sqrt{g_{11}}}}(v^2-1)={e\over{m_e}}{Q\over{r^2}}
\end{equation}
which yields
\begin{equation}
Q\simeq{{GMm_p}\over e}{L\over{L_E}}\,.
\end{equation}


\begin{references}

\reference{ch74}Carr, B.J., \& Hawking, S.W. 1974, \mnras, 168, 359
\reference{gi75}Gibbons, G.W. 1975, Commun. math. Phys, 44, 245
\reference{ha74}Hawking, S.W. 1974, \nat, 248, 30
\reference{mrt74}Maraschi, L., Reina, C., \& Treves, A. 1974, \aap, 35, 389
\reference{mrt78}Maraschi, L., Reina, C., \& Treves, A. 1978, \aap, 66, 99
\reference{mi72}Michel, F.C. 1972, \apss, 15, 153
\reference{me75}M\`esz\`aros, P. 1975, \aap, 44, 59
\reference{ntz91}Nobili, L., Turolla, R., \& Zampieri, L. 1991, \apj, 383, 250
\reference{nf89}Novikov, I.D., \& Frolov, V.P. 1989, Physics of Black Holes
(Dordrecht: Kluvers)
\reference{ru73}Ruffini, R. 1973, in Black Holes, C. DeWitt \& B.S. DeWitt eds.
(New York: Gordon and Breach)
\reference{sh73a}Shapiro, S.L. 1973a, \apj, 180, 531
\reference{sh73b}Shapiro, S.L. 1973b, \apj, 185, 69
\reference{sh70}Shvartsman, V.F. 1970, Astrofizika, 6, 309
\reference{spi56}Spitzer, L. 1956, Physics of Fully Ionized Gases (New
York: Interscience)
\reference{tu97}Turolla, R., Zane, S., Treves, A., \& Illarionov, A. 1997,
\apj, 482, 377
\reference{ztt98}Zane, S., Turolla, R., \& Treves, A. 1998, \apj, 501, 252
\reference{wa84}Wald, R.M. 1984, General Relativity (Chicago: University
of Chicago Press)

\end{references}
\end{document}